\documentclass[runningheads,citeauthoryear]{apinv}
\usepackage{epsfig,cite,graphics}
\usepackage[T2A]{fontenc}
\usepackage[cp1251]{inputenc}
\usepackage{longtable}

\begin{document}

\title{A deep decrease event in the brightness of the PMS star V350 Cep}
\titlerunning{A deep decrease event in the brightness of the PMS star V350 Cep}
\author{Evgeni H. Semkov\inst{1}, Sunay I. Ibryamov\inst{1,2}, Stoyanka P. Peneva\inst{1}}
\authorrunning{E. Semkov et al.}
\tocauthor{Evgeni Semkov}
\institute{Institute of Astronomy and National Astronomical Observatory, Bulgarian Academy of Sciences, Sofia, Bulgaria
         \and Department of Physics and Astronomy, Faculty of Natural Sciences, University of Shumen, Shumen, Bulgaria
        \newline
        \email{esemkov@astro.bas.bg}
}
\papertype{Research report. Accepted on xx.xx.xxxx}
\maketitle

\begin{abstract}
New photometric data from CCD $UBVRI$ observations of the PMS star V350 Cep during the period from March 2014 to May 2017 are presented. 
In the period April-May 2016 we registered a deep fades event in the brightness of the star with amplitudes $\Delta I$ = 1.75 mag, $\Delta R$ = 1.69 mag, $\Delta V$ = 1.77 mag and $\Delta B$ = 2.16 mag. 
Simultaneously with the fades in the brightness, the change in the star's color indices has been observed.
V350 Cep indicates the typical for stars of UXor type "blueing effect" during the deep minimum of brightness.
During the second half of 2016 V350 Cep restores its brightness to a level close to the maximum.
Since the star has been studied as a possible FUor object in previous studies, the possible cause of the deep decline is a decrease in the accretion rate.
Another possible cause is obscuration from clumps of dust orbiting at the vicinity of the star.
\end{abstract}


\section{Introduction}

The photometric variability is a widespread phenomenon during the pre-main sequence (PMS) phase of evolution of low mass stars.
In some cases, as for example, variables from the type of FU Orionis, EX Lupus, UX Orionis, some classical T Tauri stars and other young stellar objects, the registered changes in the brightness reach up to 2-5 magnitudes.
Exploring this variability gives us valuable information about processes occurring in the stars and their circumstellar neighborhood.

V350 Cep is a PMS star, which is located in the field of the reflection nebula NGC 7129. 
The region of NGC 7129 is a part of a larger structure, called the Cepheus Bubble (Kun et al. 1987), and its represents a region with active star formation (Magakian \& Movsesian 1997, Kun et al. 2008, Kun et al. 2009, Dahm \& Hillenbrand 2015). The distance to NGC 7129 is determined by Strai\v{z}ys et al. (2014) is 1.15 kpc. 
The age of the NGC 7129 star formation region determined by Straiz\v{y}s et al. (2014) is 3 Myr.

Variability of V350 Cep was discovered by Gyulbudaghian \& Sarkissian (1977) who compared their photographic plate observations in the region of NGC 7129 with the Palomar Observatory Sky Survey (POSS) plates. The star was not detected on the POSS O-plate obtained in 1954 (limit $\sim$ 21 mag) and is barely noticeable over the limit of the POSS E-plate. The measured by Gyulbudaghian \& Sarkissian (1977) brightness of V350 Cep in 1977 was approximately B = 17.50 mag. and V = 16.50 mag. The spectral class of the star was defined as M2 by Cohen \& Fuller (1985) and as M0 by Kun et al. (2009).

Following photometric observations show that the star retains its maximum brightness over the past 40 years (see Ibryamov et al. 2014 and references therein).
The observed photometric behavior of the star during the period of maximum light is typical for classical T Tauri (CTT) stars, but the historical $B/pg$-light curve is much similar to eruptive PMS stars from FU Orionis type.
The spectral observations of V350 Cep also argue about the CTT nature of the star (Magakian et al. 1999). 
A typical emission line spectrum and spectral variability characteristic of CTT stars are reported in the paper.

\section{Observations}

Presented $UBVRI$ photometric observations of V350 Cep were collected in the period from March 2014 to May 2017. 
The paper is continuation of our long-term multicolour photometric study of V350 Cep (Semkov 1993, 1996, 2002, 2004, Semkov et al. 1999, Ibryamov et al. 2014).
All observations were obtained with four telescopes, in two observatories $-$ the 2-m Ritchey-Chr\'{e}tien-Coud\'{e} (RCC), the 50/70-cm Schmidt and the 60-cm Cassegrain telescopes of the Rozhen National Astronomical Observatory (Bulgaria) and the 1.3-m Ritchey-Chr\'{e}tien (RC) telescope of the Skinakas Observatory\footnote{Skinakas Observatory is a collaborative project of the University of Crete, the Foundation for Research and Technology - Hellas, and the Max-Planck-Institut f\"{u}r Extraterrestrische Physik.} of the Institute of Astronomy, University of Crete (Greece).

The observations were performed with four different types of CCD cameras: VersArray 1300B at the 2-m RCC telescope, ANDOR DZ436-BV at the 1.3-m RC telescope, FLI PL16803 at the 50/70-cm Schmidt telescope, and FLI PL09000 at the 60-cm Cassegrain telescope. The technical parameters for the CCD cameras used, observational procedure, and data reduction process are described in Ibryamov et al. (2014). All frames were taken through a standard Johnson-Cousins set of filters. As a reference the $UBVRI$ comparison sequence reported in Semkov (2002) was used.

\begin{longtable}{lllllllll}
\caption{Photometric CCD observations of V2493 Cyg}\\
\hline\hline
\noalign{\smallskip}  
Date \hspace{1cm} &	J.D. (24...) \hspace{1mm}	&	I	\hspace{8mm} & R \hspace{6mm} & V \hspace{6mm} & B \hspace{6mm} & U \hspace{6mm} & Telescope & CCD	\\
\noalign{\smallskip}  
\hline
\endfirsthead
\caption{continued.}\\
\hline\hline
\noalign{\smallskip}  
Date \hspace{1cm} &	J.D. (24...) \hspace{1mm}	&	I	\hspace{8mm} & R \hspace{6mm} & V \hspace{6mm} & B \hspace{6mm} & U \hspace{6mm} & Telescope & CCD	\\
\noalign{\smallskip}  
\hline
\noalign{\smallskip}  
\endhead
\hline
\endfoot
\noalign{\smallskip}
21.03.2014	&	56738.483	&	14.03	&	15.27	&	16.13	&	17.25	&	$-$	&	Sch	&	FLI	\\
22.05.2014	&	56799.515	&	14.02	&	15.33	&	16.17	&	17.24	&	$-$	&	Sch	&	FLI	\\
23.05.2014	&	56801.386	&	14.11	&	15.13	&	16.12	&	17.27	&	$-$	&	2-m	&	VA	\\
23.06.2014	&	56832.363	&	13.98	&	15.04	&	16.05	&	17.25	&	$-$	&	2-m	&	VA	\\
25.06.2014	&	56834.449	&	14.02	&	15.03	&	16.01	&	17.16	&	$-$	&	2-m	&	VA	\\
28.06.2014	&	56837.443	&	13.94	&	15.11	&	15.95	&	16.99	&	$-$	&	Sch	&	FLI	\\
29.06.2014	&	56838.421	&	13.91	&	15.10	&	15.96	&	17.08	&	$-$	&	Sch	&	FLI	\\
20.07.2014	&	56859.412	&	14.04	&	15.17	&	16.03	&   $-$	&	$-$	&	60-cm	&	FLI	\\
21.07.2014	&	56860.413	&	14.16	&	15.47	&	$-$  	&   $-$	&	$-$	&	60-cm	&	FLI	\\
03.08.2014	&	56873.328	&	14.01	&	15.21	&	16.09	&	17.12	&	$-$	&	Sch	&	FLI	\\
04.08.2014	&	56874.351	&	14.00	&	15.20	&	16.01	&	17.16	&	$-$	&	Sch	&	FLI	\\
18.08.2014	&	56888.381	&	14.03	&	15.23	&	16.07	&	17.21	&	$-$	&	Sch	&	FLI	\\
19.08.2014	&	56889.318	&	14.01	&	15.15	&	16.15	&	17.30	&	$-$	&	Sch	&	FLI	\\
29.08.2014	&	56899.350	&	14.01	&	15.18	&	16.12	&	17.21	&	$-$	&	1.3-m	&	ANDOR	\\
23.09.2014	&	56924.325	&	14.09	&	15.35	&	16.33	&	  $-$	&	$-$	&	Sch	&	FLI	\\
26.11.2014	&	56973.257	&	14.02	&	15.20	&	16.17	&	17.29	&	$-$	&	Sch	&	FLI	\\
13.12.2014	&	57005.280	&	14.06	&	15.33	&	16.29	&   $-$	&	$-$	&	Sch	&	FLI	\\
14.12.2014	&	57006.340	&	14.07	&	15.34	&	16.31	&   $-$	&	$-$	&	Sch	&	FLI	\\
24.12.2014	&	57016.309	&	14.20	&	15.20	&	16.26	&	17.41	&	17.12	&	2-m	&	VA	\\
25.12.2014	&	57017.258	&	14.12	&	15.14	&	16.04	&	17.29	&	16.95	&	2-m	&	VA	\\
21.02.2015	&	57074.555	&	14.10	&	15.35	&	16.22	&	17.41	&	$-$	&	Sch	&	FLI	\\
24.04.2015	&	57136.578	&	14.04	&	15.28	&	16.16	&	17.42	&	$-$	&	Sch	&	FLI	\\
26.04.2015	&	57138.548	&	14.10	&	15.36	&	16.24	&	17.31	&	$-$	&	Sch	&	FLI	\\
20.05.2015	&	57162.507	&	14.10	&	15.38	&	16.33	&	17.49	&	$-$	&	Sch	&	FLI	\\
22.05.2015	&	57164.500	&	14.05	&	15.29	&	16.20	&	17.25	&	$-$	&	Sch	&	FLI	\\
13.06.2015	&	57186.505	&	14.13	&	15.37	&	16.25	&	17.39	&	$-$	&	Sch	&	FLI	\\
16.06.2015	&	57190.451	&	14.04	&	15.13	&	16.11	&	17.23	&	16.79	&	2-m	&	VA	\\
16.07.2015	&	57220.423	&	14.02	&	15.18	&	16.05	&	17.14	&	$-$	&	Sch	&	FLI	\\
17.07.2015	&	57221.478	&	13.99	&	15.15	&	16.03	&	17.18	&	16.87	&	Sch	&	FLI	\\
20.07.2015	&	57224.463	&	14.11	&	15.19	&	16.16	&	17.34	&	16.88	&	2-m	&	VA	\\
11.08.2015	&	57246.444	&	14.03	&	15.24	&	16.14	&	17.27	&	17.08	&	1.3-m	&	ANDOR	\\
17.08.2015	&	57252.405	&	14.11	&	15.19	&	16.21	&	17.38	&	$-$	&	2-m	&	VA	\\
24.08.2015	&	57259.393	&	14.05	&	15.28	&	16.14	&	17.28	&	$-$	&	Sch	&	FLI	\\
25.08.2015	&	57260.384	&	14.01	&	15.18	&	16.09	&	17.21	&	$-$	&	Sch	&	FLI	\\
03.09.2015	&	57269.381	&	13.99	&	15.16	&	16.02	&	17.13	&	$-$	&	Sch	&	FLI	\\
04.09.2015	&	57270.348	&	14.06	&	15.09	&	16.07	&	17.21	&	$-$	&	2-m	&	VA	\\
05.09.2015	&	57271.341	&	14.12	&	15.20	&	16.23	&	17.40	&	$-$	&	2-m	&	VA	\\
06.09.2015	&	57272.314	&	14.06	&	15.12	&	16.15	&	17.23	&	16.82	&	2-m	&	VA	\\
03.11.2015	&	57330.284	&	14.02	&	15.24	&	16.07	&	17.18	&	$-$	&	Sch	&	FLI	\\
04.11.2015	&	57331.297	&	14.02	&	15.29	&	16.17	&	17.30	&	16.98	&	Sch	&	FLI	\\
05.11.2015	&	57332.286	&	14.05	&	15.34	&	16.27	&	17.41	&	17.19	&	Sch	&	FLI	\\
06.11.2015	&	57333.286	&	14.04	&	15.29	&	16.15	&	17.28	&	$-$	&	Sch	&	FLI	\\
07.11.2015	&	57334.267	&	14.12	&	15.40	&	16.36	&	17.50	&	$-$	&	Sch	&	FLI	\\
13.12.2015	&	57370.288	&	14.05	&	15.02	&	16.00	&	17.27	&	17.54	&	2-m	&	VA	\\
14.12.2015	&	57371.281	&	14.05	&	15.05	&	16.09	&	17.30	&	16.80	&	2-m	&	VA	\\
15.12.2015	&	57372.270	&	14.06	&	15.38	&	16.33	&	17.39	&	$-$	&	Sch	&	FLI	\\
06.02.2016	&	57425.240	&	13.99	&	15.21	&	16.07	&	17.07	&	$-$	&	Sch	&	FLI	\\
04.04.2016	&	57483.434	&	15.25	&	16.21	&	17.37	&	18.67	&	$-$	&	2-m	&	VA	\\
05.04.2016	&	57484.453	&	15.11	&	16.22	&	17.26	&	18.48	&	$-$	&	2-m	&	VA	\\
06.04.2016	&	57485.448	&	15.03	&	16.22	&	17.10	&	18.43	&	$-$	&	Sch	&	FLI	\\
28.04.2016	&	57506.510	&	15.44	&	16.81	&	17.69	&	19.11	&	$-$	&	Sch	&	FLI	\\
13.05.2016	&	57522.484	&	15.74	&	16.87	&	17.84	&   $-$	&	$-$	&	Sch	&	FLI	\\
14.05.2016	&	57523.470	&	15.60	&	16.90	&	17.81	&	19.23	&	$-$	&	Sch	&	FLI	\\
31.05.2016	& 57540.470	& 14.45	& 15.57	& 16.76	& 18.27	& $-$ &	2-m	& VA \\
11.07.2016	&	57581.450	&	14.26	&	15.52	&	16.41	&	17.62	&	17.59	&	Sch	&	FLI	\\
12.07.2016	&	57582.483	&	14.32	&	15.58	&	16.62	&	17.90	&	$-$	&	Sch	&	FLI	\\
13.07.2016	&	57583.470	&	14.37	&	15.69	&	16.72	&	17.73	&	$-$	&	Sch	&	FLI	\\
02.08.2016	&	57603.437	&	14.25	&	15.34	&	16.42	&	17.78	&	17.68	&	2-m	&	VA	\\
04.08.2016	&	57605.435	&	14.18	&	15.51	&	16.46	&	17.66	&	$-$	&	Sch	&	FLI	\\
06.08.2016	&	57607.414	&	14.16	&	15.42	&	16.39	&	17.55	&	$-$	&	Sch	&	FLI	\\
02.10.2016	&	57664.349	&	14.09	&	15.34	&	16.26	&	17.46	&	16.89	&	Sch	&	FLI	\\
05.11.2016	&	57704.301	&	14.15	&	15.41	&	16.29	&	17.56	&	$-$	&	Sch	&	FLI	\\
21.11.2016  & 57714.347 & 13.96 & 14.95 & 15.93 & 17.15 & 16.84 & 2-m & VA\\
22.11.2016  & 57715.323 & 14.03 & 15.10 & 16.17 & 17.28 & $-$   & 2-m & VA\\
23.11.2016  & 57716.335 & 14.04 & 15.04 & 16.02 & 17.17 & 16.64 & 2-m & VA\\
01.01.2017  & 57755.254 & 14.05 & 15.31 & 16.22 & 17.36 & $-$   & Sch & FLI\\
02.01.2017  & 57756.281 & 14.02 & 15.31 & 16.18 & 17.43 & 17.19 & Sch & FLI\\
28.01.2017  & 57782.279 & 14.00 & 14.92 & 16.00 & 17.16 & 16.48 & 2-m & VA\\
01.02.2017  & 57786.266 & 14.02 & 15.04 & 16.12 & 17.23 & $-$   & 2-m & VA\\
15.02.2017  & 57800.224 & 13.94 & 15.15 & 16.10 & 17.20 & $-$   & Sch & FLI\\
17.02.2017  & 57801.584 & 13.86 & 15.01 & 15.86 & 16.96 & $-$   & Sch & FLI\\
04.03.2017  & 57817.499 & 13.88 & 15.03 & 15.83 & 17.01 & $-$   & Sch & FLI\\
01.04.2017  & 57845.493 & 13.78 & 14.92 & 15.72 & 16.88 & $-$   & Sch & FLI\\
03.04.2017  & 57846.518 & 13.73 & 14.82 & 15.63 & 16.72 & $-$   & Sch & FLI\\
01.05.2017  & 57875.396 & 13.93 & 14.87 & 15.85 & 16.90 & $-$   & 2-m & VA\\
18.05.2017  & 57892.473 & 13.88 & 15.07 & 15.93 & 17.03 & $-$   & Sch & FLI\\
19.05.2017  & 57893.497 & 13.91 & 14.84 & 15.71 & 16.91 & $-$   & 2-m & VA\\
30.05.2017  & 57904.474 & 13.93 & 15.10 & 15.97 & 17.13 & $-$   & Sch & FLI\\
\end{longtable}

\section{Results and discussion}

The recent photometric results of V350 Cep are summarized in Table 1. The columns of the table contains date (DD.MM.YYYY format) and Julian data (J.D.) of the observations, $UBVRI$ magnitudes of V350 Cep, telescope and CCD camera used. The typical values of the errors in the reported magnitudes are 0.01-0.02 mag for $I$- and $R$-band data, 0.01-0.04 mag for $V$-band data, 0.02-0.05 mag for $B$-band data, and 0.04-0.08 mag for $U$-band data.
The $UBVRI$ lights curves of the star during the period August 2015 $-$ November 2016 are shown in Fig. 1. 
On the figure circles denote the photometric data performed with the 2-m RCC telescope and triangles - the photometric data performed with the 50/70-cm Schmidt telescope.

\begin{figure}[!htb]
  \begin{center}
   \centering{\epsfig{file=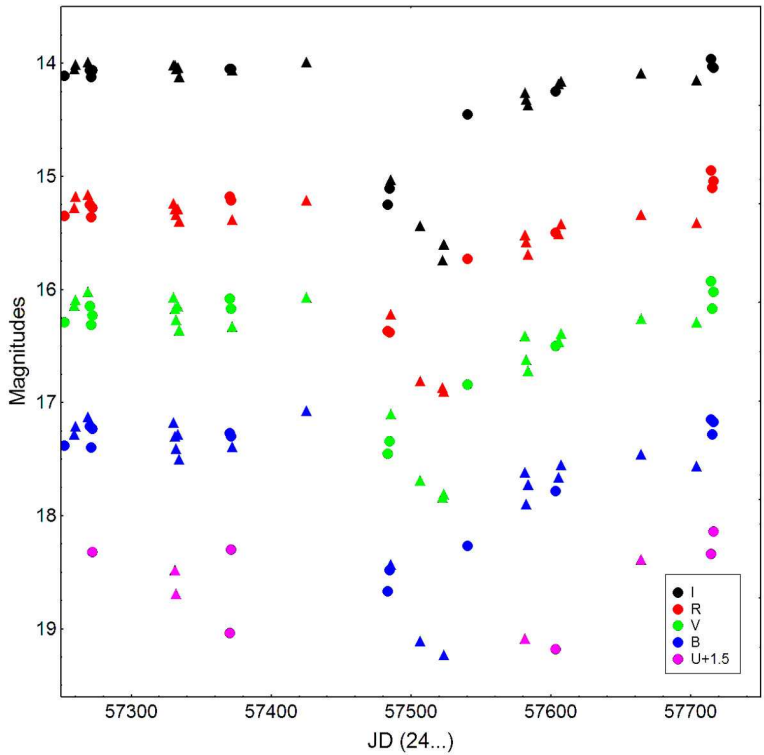, width=10cm}}
    \caption[]{$UBVRI$ light curves of V350 Cep for the period August 2015 $-$ November 2016}
    \label{countryshape}
  \end{center}
\end{figure}

\begin{figure}[!htb]
  \begin{center}
\includegraphics[width=4cm]{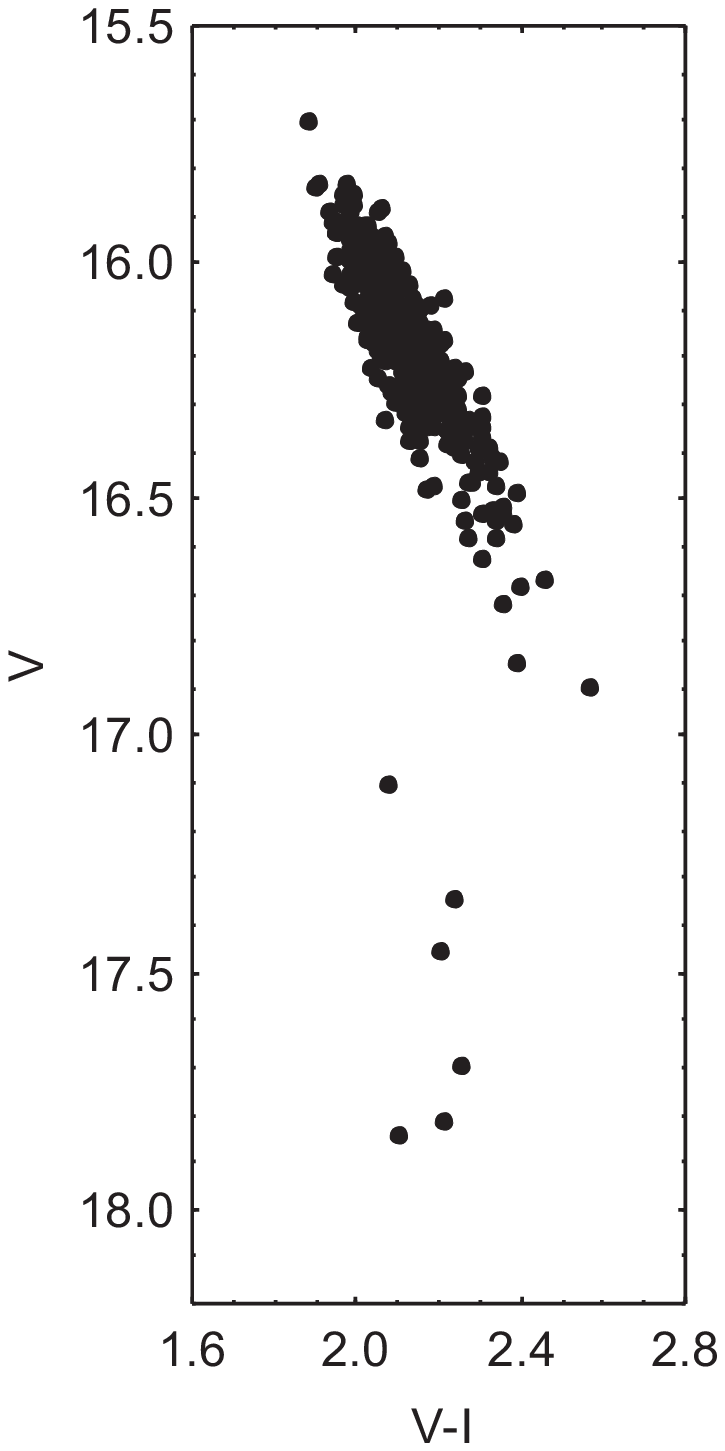}
\includegraphics[width=4cm]{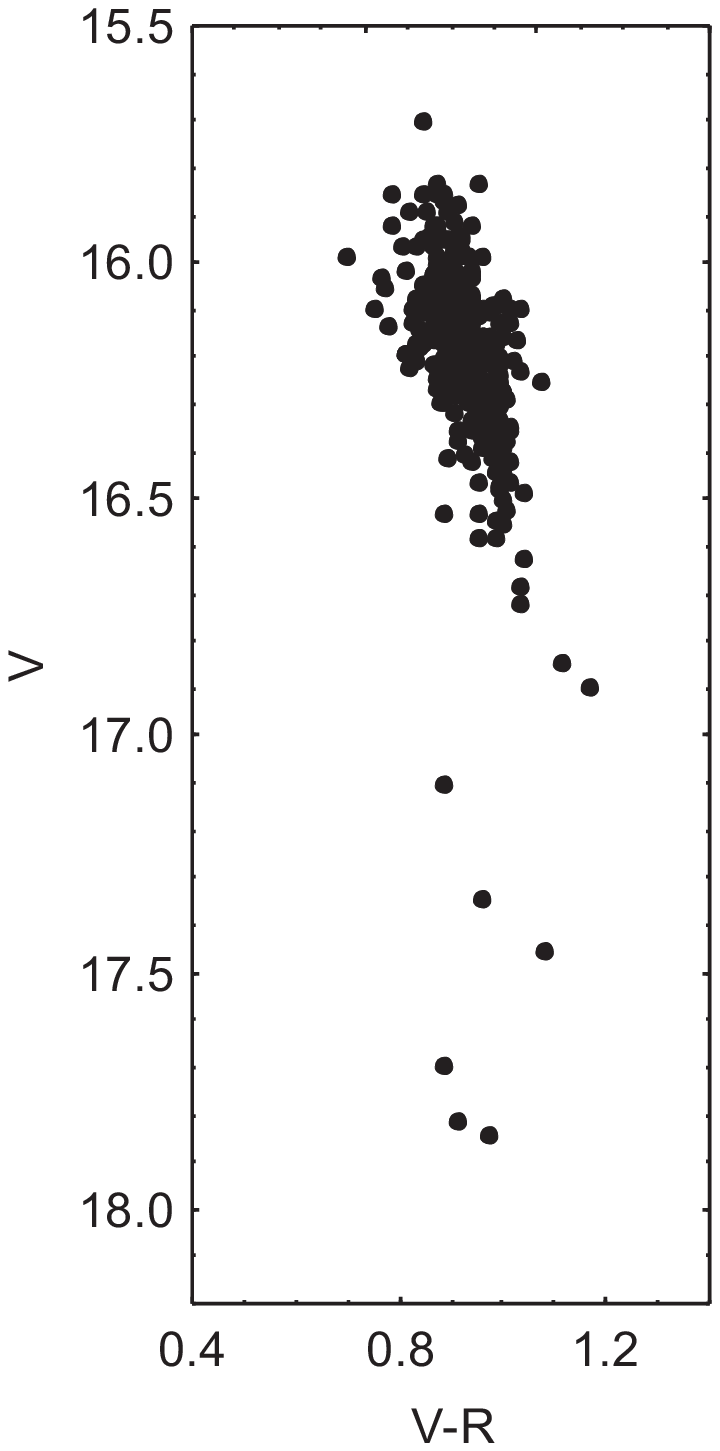}
\includegraphics[width=4cm]{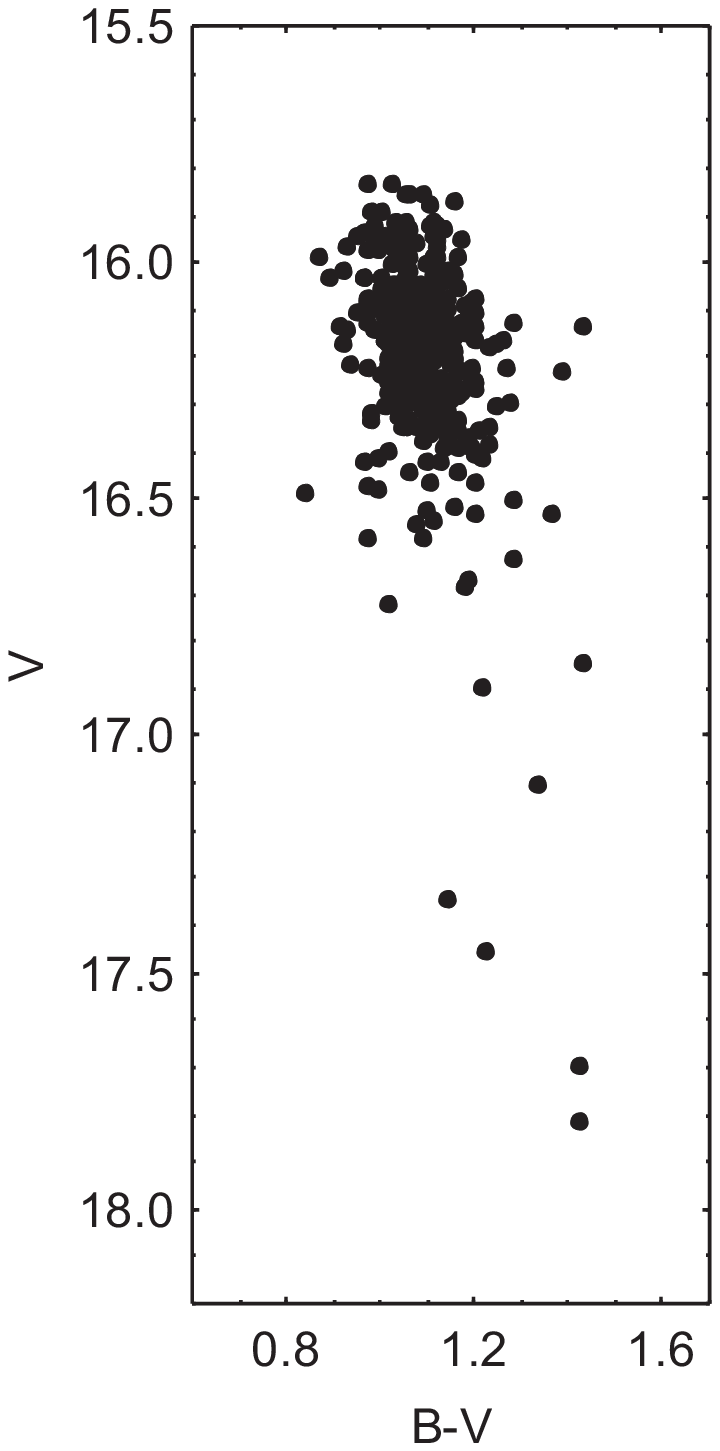}
		\caption[]{Relationship between $V$ magnitude and the $V-I$, $V-R$, and $B-V$ colour indices of V350 Cep in the period August 1994 $-$ November 2016.}
    \label{countryshape}
  \end{center}
\end{figure}

Data obtained during our long-term photometric monitoring (Semkov 1993, 1996, 2002, 2004, Semkov et al. 1999, Ibryamov et al. 2014) suggest that from 1980 until early 2016 the brightness of V350 Cep remained close to the maximum value and during the same period the star showed a low amplitude photometric variability. 
During 2016 April and May a large amplitude decrease event in the brightness of V350 Cep was observed. The registered amplitudes of the deep decline are $\Delta I$ = 1.75 mag, $\Delta R$ = 1.69 mag, $\Delta V$ = 1.77 mag and $\Delta B$ = 2.16 mag. 
From July 2016 the star regained their photometric properties, brightness at level close to the maximum, and low amplitude changes are observed.

The color variation for this period indicates a significant changes in the circumstellar environment.
The measured colour indices $V-I$, $V-R$ and $B-V$ versus stellar $V$ magnitude for the period of all our observations are plotted on Fig. 2. 
From the figure, it is seen that as usually the star becomes redder as it fades, but during the reported decrease event in 2016 April and May direction of color indexes changed, the star becomes bluer when brightness decline.
This trend is best seen for $V-I$ index.

Our observations of V350 Cep during the deep decline in the brightness indicate the presence of so-called "blueing effect", a typical feature for the PMS stars from UXor type (Bibo \& The 1990).
In accordance with the model of dust clumps obscuration, the observed colour reversal is produced by the scattered light from the small dust grains.
Normally the star becomes redder when its light is covered by dust clumps or filaments on the line of sight.
But when the obscuration rises sufficiently, the part of the scattered light in the total observed light become significant and the star color gets bluer. 
It is generally accepted that the origin of observed drops in the brightness of PMS stars and the blueing effect are due to variations of the column density of dust in the line of sight to the star.

During the PMS evolution of the stars a high amplitude of brightness variability is observed in many cases and it is difficult to separate the phenomena of eclipses and eruptions.
The large amplitude variability may result from the superposition of both phenomena, the variable accretion rate and time variable extinction, and it is very difficult to distinguish the two phenomena using only photometric data (Semkov \& Peneva 2012, Semkov et al. 2013).
In recent studies, such a scenario is used to explain the light variability of two PMS objects -- V1647 Ori (Aspin et al. 2009, Aspin 2011) and V2492 Cyg (Hillenbrand et al. 2013, K{\'o}sp{\'a}l et al. 2013). 
It seems that the time variable extinction is characteristic not only of UXor variables but it is a common phenomenon during the evolution of all types of PMS stars.

\begin{figure}[!htb]
  \begin{center}
   \centering{\epsfig{file=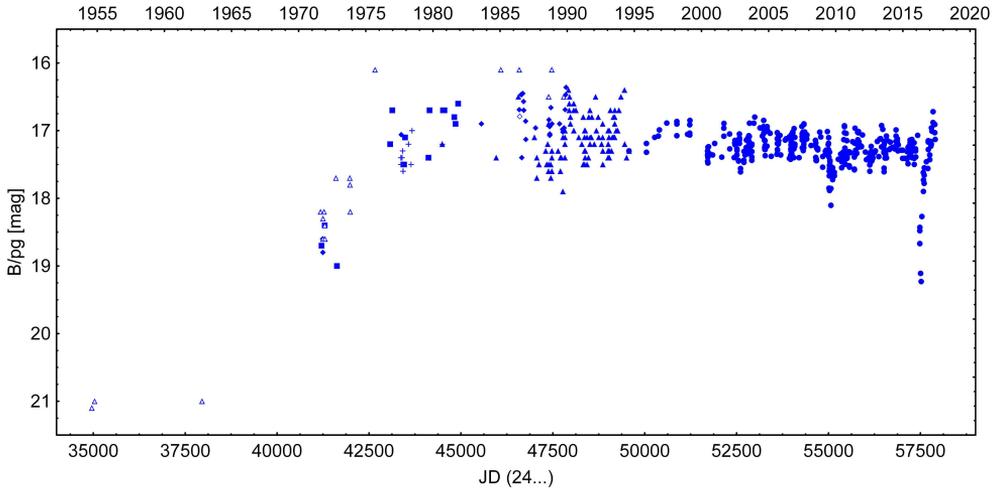, width=13cm}}
    \caption[]{$B/pg$-light curve of V350 Cep during the period 1950--2017.}
    \label{countryshape}
  \end{center}
\end{figure}

The long-term $B/pg$-light curve of V350 Cep from all available published observations are shown in Fig. 3. 
The meanings of different symbols used are as in Ibryamov et al. (2014). 
It is seen from the figure that the historical period of strong increase in brightness of V350 Cep continue to about 1978 and then begin a period of irregular variability around the level of maximum brightness lasting to 2016. 
However during the years, a slight tendency of dropping in brightness is observed, which is an indication of eventual membership to the group of the PMS eruptive stars from FUor type.

\section{Conclusion}
Future observations of V350 Cep are very important for clarifying the nature of the star.
Only the long-term multicolor photometric and spectral observations can determine the nature of the observed phenomena.
We plan to continue our observation of the star, both photometric and spectral during the coming years.

{\it Acknowledgments:} This research was partially supported by the Bulgarian National Science Fund of the Ministry of Education and Science under grants DN 08-1/2016 and DM 08-2/2016 and by funds of the project RD-08-102 of Shumen University.
The authors thank the Director of Skinakas Observatory Prof. I. Papamastorakis and Prof. I. Papadakis for the award of telescope time. 
The research has made use of the NASA's Astrophysics Data System Abstract Service.

\end{document}